\documentclass[12pt]{article}
\usepackage{epsfig}
\usepackage{amssymb}
\usepackage{amsmath}
\usepackage{amsfonts}
\usepackage{graphicx}
\usepackage{mathrsfs}
\usepackage[dvips]{color}
\usepackage{multirow}

\newcommand{\R}{\mathbb{R}}

\newcommand{\fc}{\mathfrak{c}}

\newcommand{\ff}{\mathfrak{f}}

\newcommand{\fz}{\mathfrak{z}}

\newcommand{\bfa}{\mathbf{a}}

\newcommand{\bfe}{\mathbf{e}}

\newcommand{\bk}{\mathbf{k}}

\newcommand{\bp}{\mathbf{p}}
\newcommand{\bx}{\mathbf{x}}
\newcommand{\bu}{\mathbf{u}}

\newcommand{\bA}{\mathbf{A}}

\newcommand{\bM}{\mathbf{M}}

\newcommand{\be}{\begin{equation}}
\newcommand{\ee}{\end{equation}}
\newcommand{\bea}{\begin{eqnarray}}
\newcommand{\eea}{\end{eqnarray}}
\newcommand{\nn}{\nonumber}
\newcommand{\kt}{\rangle}
\newcommand{\br}{\langle}

\newcommand{\ed}{\end{document}}

\newcommand{\bi}{\begin{itemize}}
\newcommand{\ei}{\end{itemize}}

\newcommand{\bce}{\begin{center}}
\newcommand{\ece}{\end{center}}

\newcommand{\sD}{\mathscr{D}}

\newcommand{\sF}{\mathscr{F}}

\newcommand{\bzero}{{\boldsymbol{0}}}

\newcommand{\for}{{\mbox{\rm for}}}

\begin{document}

%\title{Pseudo-Hermitian Hamiltonians in real potential scattering and their exceptional points}

\title{Renormalization of multi-delta-function point scatterers in two and three dimensions, the coincidence-limit\\ problem, and its resolution}

\author{Farhang Loran\thanks{E-mail address: loran@iut.ac.ir}~ and
Ali~Mostafazadeh\thanks{E-mail address:
amostafazadeh@ku.edu.tr}\\[6pt]
$^{*}$Department of Physics, Isfahan University of Technology, \\ Isfahan 84156-83111, Iran\\[6pt]
$^\dagger$Departments of Mathematics and Physics, Ko\c{c}
University,\\  34450 Sar{\i}yer, Istanbul, T\"urkiye}

\date{ }
\maketitle

\begin{abstract}
In two and three dimensions, the standard treatment of the scattering problem for a multi-delta-function potential, $v(\mathbf{r})=\sum_{n=1}^N\mathfrak{z}_n\delta(\mathbf{r}-\mathbf{a}_n)$, leads to divergent terms. Regularization of these terms and renormalization of the coupling constants $\mathfrak{z}_n$ give rise to a finite expression for the scattering amplitude of this potential, but this expression has an important short-coming; in the limit where the centers $\mathbf{a}_n$ of the delta functions coincide, it does not reproduce the formula for the scattering amplitude of a single-delta-function potential, i.e., it seems to have a wrong coincidence limit. We provide a critical assessment of the standard treatment of these potentials and offer a resolution of its coincidence-limit problem. This reveals some previously unnoticed features of this treatment. For example, it turns out that the standard treatment is incapable of determining the dependence of the scattering amplitude on the distances between the centers of the delta functions. This is in sharp contrast to the treatment of this problem offered by a recently proposed dynamical formulation of stationary scattering. For cases where the centers of the delta functions lie on a straight line, this formulation avoids singularities of the standard approach and yields an expression for the scattering amplitude which has the correct coincidence limit. 
%\vspace{2mm}

%\noindent PACS numbers: 03.65.Nk, 42.25.Bs\vspace{2mm}

%\noindent Keywords: Exceptional point, pseudo-Hermitian operator, scattering, transfer matrix, biorthonormal system
\end{abstract}

\section{Introduction}

The emergence of singularities in the treatment of delta-function potentials in two and three dimensions and the utility of various regularization/renormalization schemes for their removal have been known for over four decades \cite{thorn,jackiw,mead,manuel,Adhikari1,Adhikari2,Mitra,Rajeev-1999,Nyeo,Camblong,teo-2006,teo-2010,teo-2013,ferkous,ap-2019}. These schemes allow for the solution of the scattering problem for the multi-delta-function potentials,
	\be
	v(\bx)=\sum_{j=1}^N\fz_j\delta(\bx-\bfa_j),
	\label{multi-delta}
	\ee
where $\bx$ is the position vector, $\fz_j$ are real or complex coupling constants, and $\bfa_j$ are the centers of the delta functions. In two dimensions, they lead to the following formula for the scattering amplitude \cite{ap-2019}.\footnote{We offer a derivation of this formula in Sec.~\ref{S2}.}
	\be
	\ff(\bk',\bk)=-\sqrt{\frac{i}{8\pi k}}\sum_{m,n=1}^N A_{mn}^{-1}\, e^{i(\bfa_n\cdot\bk-\bfa_m\cdot\bk')}.
	\label{f=N}
	\ee
Here $\bk$ is the incident wave vector, $k:=|\bk|$, $\bk':=k \bx/r$ is the wave vector for the scattered wave, $r:=|\bx|$, $A_{mn}^{-1}$ are the entries of the inverse of the matrix $\bA:=[A_{mn}]$ with entries
	\be
	A_{mn}:=\left\{\begin{array}{ccc}
	\tilde\fz_m^{-1}+\frac{i}{4}&\for& m=n,\\[3pt]
	\frac{i}{4}H_0^{(1)}(k|\bfa_m-\bfa_n|)&\for&m\neq n,\end{array}\right.
	\label{A=}
	\ee
$\tilde\fz_m$ are the renormalized coupling constants, $H_0^{(1)}$ is the zero-order Hankel function of the first kind, and we use units and conventions where the time-independent Schr\"odinger equation takes the form,
	\be
	\left[-\nabla^2+v(\bx)\right]\psi(\bx)=k^2\psi(\bx),
	\label{sch-eq}
	\ee
and its scattering solutions satisfy,
	\be
	\psi(\bx)\to\frac{1}{(2\pi)^{d/2}}\left[e^{i\bk\cdot\bx}+\frac{\ff(\bk',\bk)\,e^{ikr}}{r^{\frac{d-1}{2}}}\right]
	~~\for~~r\to\infty,
	\label{asym-2D}
	\ee
in $d$ dimensions.

For $N=1$, (\ref{f=N}) and (\ref{A=})	give
	\be
	\ff(\bk',\bk)=-\sqrt{\frac{i}{8\pi k}}\:\frac{e^{i\bfa_1\cdot(\bk-\bk')}}{\tilde\fz_1^{-1}+i/4}.
	\label{f=1}
	\ee
For $N\geq 2$ and $m\neq n$, $A_{mn}$ diverges in the coincidence limit, $\bfa_m\to\bfa_n$. Because $\bA$ is a symmetric matrix, this implies that the entries of $\bA^{-1}$ and, in view of (\ref{f=N}), the scattering amplitude $\ff(\bk',\bk)$ tend to zero in this limit. This in particular means that if the distance between the centers of any two of the delta functions in (\ref{multi-delta}) becomes much smaller than $k^{-1}$, the scattering amplitude (\ref{f=N}) diminishes in magnitude regardless of the number, position, and value of the coupling constants for the remaining $N-2$ delta functions. In particular, according to (\ref{f=N}) and (\ref{A=}), the multi-delta-function potential (\ref{multi-delta}) seizes to scatter waves, if 
	\be
	\bfa_n\to\bfa_1~~~\mbox{for all}~~~n\in\{2,3,\cdots,N\}.
	\label{CL}
	\ee
But this contradicts the fact that in this limit (\ref{multi-delta}) turns into a single-delta-function potential with coupling constant $\fz:=\sum_{n=1}^N\fz_n$, and consequently its scattering amplitude must be given by (\ref{f=1}) with $\tilde\fz_1$ changed to $\tilde\fz$. The same difficulties arise in dealing with  
multi-delta-function potentials in three dimensions. The purpose of this article is to provide a comprehensive treatment of the scattering problem for multi-delta-function potentials in two and three dimensions that offers a resolution of these difficulties. 

The main motivation for the present work is provided by a recently-developed dynamical formulation of stationary scattering (DFSS) \cite{pra-2016,pra-2021} which turns out to offer a singularity-free treatment of the scattering problem for (\ref{multi-delta}) in two dimensions when the centers of the delta functions lie on a straight line \cite{jpa-2018}. For this configuration, it leads to (\ref{f=N}) with $A_{mn}$ given by
	\be
	A_{mn}:=\left\{\begin{array}{ccc}
	\fz_m^{-1}+\frac{i}{4}&\for& m=n,\\[3pt]
	\frac{i}{4}J_0^{(1)}(k|\bfa_m-\bfa_n|)&\for&m\neq n,\end{array}\right.
	\label{A=2}
	\ee
where $J_0$ is the zero-order Bessel function of the first kind. For $N=1$, this agrees with (\ref{f=1}), if we identify $\tilde\fz_1$ with $\fz_1$;
	\be
	\ff(\bk',\bk)=-\sqrt{\frac{i}{8\pi k}}\:\frac{e^{i\bfa_1\cdot(\bk-\bk')}}{\fz_1^{-1}+i/4}.
	\label{f=1-DFSS}
	\ee
For $N\geq 2$, substituting (\ref{A=2}) in (\ref{f=N}), we find an expression for the scattering amplitude that, as we show in Appendix~A, has the correct coincidence limit. 

%Recently, we have traced back the implicit regularization property of DFSS for a single-delta-function potential to the action of a projection operator that has the effect of putting a momentum cut-off on the domain of the same type of singular integrals that appear in the traditional treatment of this potential \cite{pra-2021}. We therefore begin our analysis by offering a review of the standard treatment of multi-delta-function potentials in two and three dimensions that involves a cut-off renormalization of these potentials. 

%The organization of this article is as follows. In Sec.~\ref{S2}, we review the cut-off renormalization of the multi-delta-function potentials in two and three dimensions using the standard approach based on the  Lippmann-Schwinger equation. In Sec.~\ref{S3}, we elucidate the conceptual roots of the coincidence-limit problem and use the results obtain within the framework of DFSS to outline a comprehensive solution for this problem. In Sec~\ref{S4}, we present a summary of our findings and our concluding remarks.

\section{Standard treatment of multi-delta-function potentials}
\label{S2}

Consider the multi-delta-function potential (\ref{multi-delta}) in $d$ dimensions. The Lippmann-Schwinger equation for this potential has the form
	\be
	|\psi\kt=|\bk\kt+\hat G\hat V|\psi\kt,
	\label{LS-eq}
	\ee
where 
	\begin{align}
	%&\br \bx|\psi_0\kt:=\frac{e^{i\bk_0\cdot\bx}}{(2\pi)^{d/2}},
	%\label{incident}\\
	&\hat G:=\lim_{\epsilon\to 0^+}\frac{1}{k^2-\hat\bp^2+i\epsilon},
	&&\hat V:=\sum_{n=1}^N\fz_n|\bfa_n\kt\br\bfa_n|.
	\nn%\label{resolvent}
	\end{align}
In the position representation, (\ref{LS-eq}) takes the form,
	\be
	\br\bx|\psi\kt=\br\bx|\bk\kt+\sum_{n=1}^N\fz_n G(\bx-\bfa_n)\br\bfa_n|\psi\kt,
	\label{LS-eq2}
	\ee
where $G(\bx-\bx'):=\br\bx|\hat G|\bx'\kt$ is the outgoing Green's function for the Helmholtz operator, $\nabla^2+k^2$. It is easy to see that
	\be
	G(\bx)=
	\lim_{\epsilon\to 0^+}\int_{\R^d}
	\frac{d^d\bp}{(2\pi)^d}\:\frac{e^{i\bx\cdot\bp}}{k^2-\bp^2+i\epsilon}.
	\label{G=int}
	\ee	
	
Let us introduce $X_n:=\fz_n\br\bfa_n|\psi\kt$, and write (\ref{LS-eq2}) in the form
	\be
	\br\bx|\psi\kt=\br\bx|\bk\kt+\sum_{n=1}^N G(\bx-\bfa_n)X_n.
	\label{LS-eq3}
	\ee
For $\bx=\bfa_m$, this gives the following system of equations for $X_n$.
	\be
	\sum_{n=1}^N A_{mn} X_n=\frac{e^{i\bfa_m\cdot\bk}}{(2\pi)^{d/2}},
	\label{sys1}
	\ee
where
	\be
	A_{mn}:=\fz_n^{-1}\delta_{mn}-G(\bfa_m-\bfa_n).
	\label{Amn=1}
	\ee
Eq.~(\ref{sys1}) has a unique solution provided that the matrix $\bA$ of its coefficients $A_{mn}$ is invertible.\footnote{This happens when the Schr\"odinger opertor $-\nabla^2+v(\bx)$ has no spectral singularities \cite{prl-2009}.} In this case we can express the solution in the form, 
	\[X_n=\frac{1}{(2\pi)^{d/2}}\sum_{m=1}^NA_{nm}^{-1}\,e^{i\bfa_m\cdot\bk}.\]
Substituting this equation in (\ref{LS-eq3}), we obtain
	\be
	\br\bx|\psi\kt=\br\bx|\bk\kt+\frac{1}{(2\pi)^{d/2}}
	\sum_{m,n=1}^NG(\bx-\bfa_n)A_{nm}^{-1}\,e^{i\bfa_m\cdot\bk}.
	\label{LS-sol}
	\ee
	
In two and three dimension, we can evaluate the integral on the right-hand side of (\ref{G=int}), \cite{MF}. The result is 
	\be
	G(\bx)=\left\{\begin{array}{ccc}
	%\displaystyle-\frac{i e^{ik|\bx-\bx'|}}{2k}&\for & d=1,\\[6pt]
	\displaystyle-\frac{i}{4}\,H_0^{(1)}(kr) &\for & d=2,\\[3pt]
	\displaystyle-\frac{1}{4\pi}\displaystyle\frac{e^{ikr}}{r}&\for & d=3.\end{array}\right.
	\label{G=}
	\ee
This implies,
	\be
	G(\bx-\bx')\to\left\{\begin{array}{ccc}
	%\displaystyle-\frac{i e^{ik(r-r')}}{2k}&\for & d=1,\\[6pt]
	\displaystyle-\sqrt{\frac{i}{8\pi k r}}\;e^{-i\bk'\cdot\bx'}e^{ikr} &\for & d=2\\[6pt]
	\displaystyle-\frac{1}{4\pi}\displaystyle\frac{e^{-i\bk'\cdot\bx'}e^{ikr}}{r}&\for & d=3\end{array}\right\}~~{\rm as}~~r\to\infty.
	\label{G-asym}
	\ee
In view of (\ref{asym-2D}), (\ref{LS-sol}), and (\ref{G-asym}),
	\begin{align}
	&\ff(\bk',\bk)=\frac{c_d}{\sqrt{k^{3-d}}} \sum_{m,n=1}^N A_{mn}^{-1}\, e^{i(\bfa_n\cdot\bk-\bfa_m\cdot\bk')},
	&&c_d:=\left\{\begin{array}{ccc}
	-\sqrt{i/8\pi}&\for& d=2,\\[3pt]
	-1/4\pi &\for& d=3.\end{array}\right.
	\label{f=N-3d}
	\end{align}	

The difficulties associated with delta-function potentials in two and three dimensions stem from the fact that the right-hand side of (\ref{G=}) blows up for $\bx=\bzero$. In view of (\ref{Amn=1}), this implies $A_{nn}=\infty$. One can regularize $G(\bx)$ and perform a renormalization of the coupling constants to turn (\ref{f=N-3d}) into sensible expressions for the scattering amplitude. In the remainder of this section, we review the utility of the cut-off renormalization scheme for this purpose.

\subsection{Renormalization of multi-delta-function potentials in 2D}

For $d=2$, we can evaluate the integral on the right-hand side of (\ref{G=int}) using a coordinate system in the momentum plane in which $\bx$ points along the $x$-axis. Labeling the polar coordinates in this plane by $(p,\varphi)$ and noting that
$\int_0^{2\pi}d\varphi\: e^{ix\cos\varphi}=2\pi J_0(x)$, we have \cite{MF},
	\be
	G(\bx)=\lim_{\epsilon\to 0^+}\int_0^\infty \frac{dp}{2\pi}\:
	\frac{p\,J_0(pr)}{k^2-p^2+i\epsilon}=
	-\frac{i}{4}\,H_0^{(1)}(kr).
	\label{G=2d}
	\ee
For $r\to 0$, $J_0(pr)\to 1$, and the integral in this equation develops a logarithmic singularity. We regularize this singularity by introducing a momentum cut-off $\Lambda$. This changes $G(\bx)$ to
	\be
	G_\Lambda(\bx):=\lim_{\epsilon\to 0^+}\int_0^\Lambda \frac{dp}{2\pi}\:
	\frac{p\,J_0(pr)}{k^2-p^2+i\epsilon}.
	\label{G-Lambda}
	\ee
For $\bx=\bzero$, $r=0$, and (\ref{G-Lambda}) gives
	\bea
	G_\Lambda(\bzero)&=&-\frac{1}{4\pi}\ln\left(\frac{\Lambda^2}{k^2}-1\right)-\frac{i}{4}=
	-\frac{1}{2\pi}\ln\left(\frac{\Lambda}{k}\right)-\frac{i}{4}+O\big((k/\Lambda)^2\big),
	\label{G-mu-zero}
	\eea
where $O(x^n)$ denotes the terms of order $n$ and higher in powers of $x$. Because
	\be
	H_0^{(1)}(x)=\frac{2i}{\pi}\left(\ln\frac{x}{2}+\gamma\right)+1+O(x^2),
	\label{H-asymp}
	\ee
where $\gamma$ is the Euler number, we can use (\ref{G=2d}) and (\ref{G-mu-zero}) to show that for every positive real number $\alpha$,
	\bea
	\lim_{r\to 0}\Big[G(\bx)-G_{_{\!\mbox{\footnotesize$\frac{\alpha}{r}$}}}\!(\bzero)\Big]=\frac{\gamma+\ln(\alpha/2)}{2\pi}.
	\label{G=G+}
	\eea
In light of this relation, we can identify $G(\bx)$ with $G_{_{\!\mbox{\footnotesize$\frac{\alpha}{r}$}}}\!(\bzero)+[\gamma+\ln(\alpha/2)]/2\pi$ whenever $r:=|\bx|$ is much smaller than $\alpha/k$. 

In order to arrive at a finite (and nonzero) expression for the scattering amplitude (\ref{f=N-3d}), first we need to reinterpret the coupling constants $\fz_n$ appearing in the expression for the potential (\ref{multi-delta}) as the bare coupling constants $\mathring\fz_n$ which have no a priori physical meaning. In other words, the first step of the renormalization program is to model the scattering problem using the potential,
	\be
	\mathring v(\bx)=\sum_{j=1}^N\mathring\fz_j\delta(\bx-\bfa_j).
	\nn%\label{multi-delta-bare}
	\ee
Next, we let $\mu$ be an arbitrary reference momentum scale, set $\alpha:=2 e^{-\gamma}\mu/k$, and introduce the renormalized coupling constants,
	\be
	\tilde\fz_n:=\left(\mathring\fz_n^{-1}+\frac{1}{2\pi}\ln\frac{\Lambda}{\mu}\right)^{-1}=
	\left(\mathring\fz_n^{-1}+\frac{1}{2\pi}\ln\frac{\Lambda}{k}-\frac{\gamma+\ln(\alpha/2)}{2\pi}\right)^{-1}.
	\label{zn-renorm-1}
	\ee
If we suppose that $\mathring\fz_n$ depends on $\Lambda$ in such a way that $\tilde\fz_n$ is $\Lambda$-independent, we can use (\ref{G-mu-zero}) -- (\ref{zn-renorm-1}) to show that for $r=\alpha/\Lambda\to 0$,
	\be
	\mathring\fz_n^{-1}-G(\bx)~
	\to~  \mathring\fz_n^{-1}-G_{\Lambda}(\bzero)-\frac{\gamma+\ln(\alpha/2)}{2\pi}=
	\tilde\fz_n^{-1}+\frac{i}{4}.\nn
	\ee
In view of (\ref{Amn=1}), this suggests
	\be
	A_{nn}=\tilde\fz_n^{-1}+\frac{i}{4}.
	\label{Ann=2}
	\ee
Equation (\ref{A=}) follows from (\ref{Amn=1}), (\ref{G=}), and (\ref{Ann=2}).

\subsection{Renormalization of multi-delta-function potentials in 3D}

For $d=3$, we express the right-hand side of  (\ref{G=int}) in spherical coordinates in the momentum space and put a cut-off $\Lambda$ on the radial coordinate to obtain the regularized Green's function $G_\Lambda(\bx)$. Doing the angular integrals, we then find
	\bea
	G_\Lambda(\bx)&=&\frac{1}{2\pi^2 r}
	\lim_{\epsilon\to 0^+}\int_0^\Lambda dp\:\frac{p\sin(rp)}{k^2-p^2+i\epsilon},\\
	G_\Lambda(\bzero)&=&\lim_{r\to 0}G_\Lambda(\bx)=
	\frac{1}{2\pi^2}\lim_{\epsilon\to 0^+}\int_0^\Lambda dp\:\frac{p^2}{k^2-p^2+i\epsilon}
	=-\frac{\Lambda}{2\pi^2}-\frac{ik}{4\pi}+O(\Lambda^{-1}).
	\eea
The latter relation together with (\ref{G=}) imply $\lim_{r\to 0}\Big[G(\bx)-G_{\pi/2r}(\bzero)\Big]=0$.

We can introduce renormalized coupling constants $\tilde\fz_n$ according to 
	$\tilde\fz_n:=\left(\mathring\fz_n+ \Lambda/2\pi^2\right)^{-1}$.
Supposing that $\mathring\fz_n$ depend on $\Lambda$ in such a way that $\tilde\fz_n$ is $\Lambda$-independent, we can show that for $r=\pi/2\Lambda\to 0$,
	\be
	\mathring\fz_n^{-1}-G(\bx)\to \mathring\fz_n^{-1}-G_\Lambda(\bzero)=\tilde\fz_n^{-1}+\frac{ik}{4\pi}.
	\label{ren-zn-3d}
	\ee
Relations (\ref{Amn=1}), (\ref{G=}), and (\ref{ren-zn-3d}) lead to the following three-dimensional analog of (\ref{A=}).
	\be
	A_{mn}=\left\{\begin{array}{ccc}
	\tilde\fz_n^{-1}+\frac{ik}{4\pi}&\for&m=n,\\[3pt]
	\displaystyle\frac{e^{ik|\bfa_m-\bfa_n|}}{4\pi|\bfa_m-\bfa_n|}&\for& m\neq n.
	\end{array}\right.
	\label{Amn-3d=}	
	\ee
For $N=1$, (\ref{f=N-3d}) and (\ref{Amn-3d=}) imply
	\be
	\ff(\bk',\bk)=-\frac{e^{i\bfa_1\cdot(\bk-\bk')}}{4\pi\,\tilde\fz_1^{-1}+ik}.
	\label{f1-3d=}
	\ee
Furthermore, according to (\ref{Amn-3d=}), $A_{mn}$ diverges for $k|\bfa_m-\bfa_n|\to 0$, if $m\neq n$. In this case, $\bA^{-1}$ tends to the zero matrix, and (\ref{f=N-3d}) predicts that the multi-delta-function potential (\ref{multi-delta}) ceases to scatter waves effectively, if there is a pair of delta functions contributing to this potential such that the distance between their centers is much less than $k^{-1}$. This observation together with (\ref{f1-3d=}) underline the coincidence-limit problem associated with the use of (\ref{Amn-3d=}) in (\ref{f=N-3d}) for $d=3$.

\section{Resolution of the coincidence limit problem}
\label{S3}

The renormalization schemes used in the treatment of the multi-delta-function potentials in two and three dimensions involve subtraction of unwanted infinities. This procedure however is not unique. For example, in the cut-off renormalization of these potentials, we can define the renormalized coupling constants according to
	\be
	\tilde\fz_n:=\left\{\begin{array}{ccc}
	\left(\mathring\fz_n^{-1}+\frac{1}{2\pi}\ln\frac{\Lambda}{\mu}- \fc_{n}\right)^{-1}&\for&d=2,\\[6pt]
	\left(\mathring\fz_n^{-1}+\frac{\Lambda}{2\pi^2}- \fc_{n}\right)^{-1}&\for&d=3,
	\end{array}\right.\nn
	\ee
where $ \fc_{n}$ are arbitrary real constants. This choice only changes the diagonal entries of the matrix $\bA$, which take the form
	\be
	A_{nn}:=\left\{\begin{array}{ccc}
	\tilde\fz_n^{-1}+ \fc_{n}+\frac{i}{4}&\for&d=2,\\[6pt]
	\tilde\fz_n^{-1}+ \fc_{n}+\frac{ik}{4\pi}&\for&d=3.
	\end{array}\right.\nn
	\ee
The arbitrariness related to the subtraction of infinities may seem irrelevant to the singularities we encounter in performing the coincidence limit of the scattering amplitude $\ff(\bk',\bk)$, because the latter are related to the off-diagonal entries of $\bA$. A closer examination of the structure of $\ff(\bk',\bk)$, however, reveals a different picture. 

The renormalized couplings constant $\tilde\fz_n$ can not only depend on the wavenumber of the incident wave but on the distances between the centers of the delta functions constituting the potential, i.e., $\ell_{mn}:=|\bfa_m-\bfa_n|$. The dependence of $\tilde\fz_n$ on $\ell_{mn}$ should be such that the scattering amplitude (\ref{f=N-3d}) has the correct coincidence limit. The standard treatment of multi-delta-function potentials, which we have reviewed in the preceding section, does not determine the nature of the $\ell_{mn}$-dependence of $\tilde\fz_n$. It only provides information about the dependence of the scattering amplitude on the wave vector $\bk'$ for the scattered wave. This is clear from (\ref{f=N-3d}) particularly if we write it in the form 
	\be
	\ff(\bk',\bk)=\frac{c_d}{\sqrt{k^{3-d}}}\sum_{m=1}^N \ff_{m}(\bk) e^{-i\bfa_m\cdot\bk'},
	\label{f=N-2}
	\ee
where
	\be
	\ff_{m}(\bk):=\sum_{n=1}^N A_{mn}^{-1}\, e^{i\bfa_n\cdot\bk}.
	\label{fm=N}
	\ee
Note that according to (\ref{A=}) and (\ref{Amn-3d=}), $A_{mn}^{-1}$ and consequently $\ff_{m}(\bk)$ depend on $\bk$, $\bfa_n$, and $\tilde\fz_n$. The dependence of $\tilde\fz_n$ on $\ell_{mn}$ should be such that (\ref{f=N-2}) has the correct coincidence limit. To arrive at a more detailed description of the $\ell_{mn}$-dependence of $\tilde\fz_n$, we explore double-delta-function potentials.

\subsection{Double-delta-function potentials in 2D}

Consider a general double-delta-function potential, i.e., (\ref{multi-delta}) with $N=2$, in two dimensions. We can choose a coordinate system in which $\bfa_1=\bzero$ and $\bfa_2=\ell\,\bfe_y$, where $\ell$ is a positive real parameter, $\bfe_u$ is the unit vector pointing along the positive $u$-axis, and $u\in\{x,y\}$. In this coordinate system the double-delta-function potential reads,
	\be
	v(x,y)=\fz_1\delta(x)\delta(y)+\fz_2\delta(x)\delta(y-\ell).
	\label{double-delta}
	\ee
Letting $\theta_0$ and $\theta$ respectively denote the incidence and scattering angles, so that $\bk\cdot\bfe_x=k\cos\theta_0$ and $\bk'\cdot\bfe_x=k\cos\theta$, we can use (\ref{f=N}), (\ref{f=N-2}), and (\ref{fm=N}) to show that 
	\begin{align}
	&\ff(\bk',\bk)=-\sqrt{\frac{i}{8\pi k}}\,\Big[\ff_1(\bk)+\ff_2(\bk)e^{-ik\ell\sin\theta}\Big],
	\label{f=2-delta}\\
	&\ff_1(\bk)= A^{-1}_{11}+A^{-1}_{12}e^{ik\ell\sin\theta_0}=
	\frac{4\big[4\tilde\fz_2^{-1}+i-i H_0^{(1)}(k\ell)e^{ik\ell\sin\theta_0}\big]}{
	(4\tilde\fz_1^{-1}+i)(4\tilde\fz_2^{-1}+i)+H_0^{(1)}(k\ell)^2},
	\label{f1=}\\[3pt]
	&\ff_2(\bk)= A^{-1}_{21}+A^{-1}_{22}e^{ik\ell\sin\theta_0}=	
	\frac{4\big[(4\tilde\fz_1^{-1}+i)e^{ik\ell\sin\theta_0}-i H_0^{(1)}(k\ell)\big]}{
	(4\tilde\fz_1^{-1}+i)(4\tilde\fz_2^{-1}+i)+H_0^{(1)}(k\ell)^2}.
	\label{f2=}
	\end{align}
	
Demanding that $\ff_n$ tend to nonzero regular functions $\ff_{0n}$ for $\ell\to 0$, so that
	\be
	\ff_{0n}(\bk):=\lim_{\ell\to 0}\ff_n(\bk),
	\label{limit}
	\ee
we can use (\ref{H-asymp}), (\ref{f1=}), and (\ref{f2=}) to determine the small-$\ell$ behavior of the renormalized coupling constants $\tilde\fz_n$. This leads to the following expressions for $\tilde\fz_n$ whose derivation we present in Appendix~B.
	\begin{align}
	&\tilde\fz_1=\left\{
	-\frac{i}{4}\Big[\eta(\bk) H_0^{(1)}(k\ell)+1\Big]+\frac{1}{\ff_{01}(\bk)}+\phi_1(\bk,\ell)\right\}^{-1},
	\label{fz1=}\\
	&\tilde\fz_2=\left\{
	-\frac{i}{4}\Big[\eta(\bk)^{-1} H_0^{(1)}(k\ell)+1\Big]+\frac{1}{\ff_{02}(\bk)}+\phi_2(\bk,\ell)\right\}^{-1}.
	\label{fz2=}
	\end{align}
Here $\phi_n(\bk,\ell)$ are functions such that $\displaystyle\lim_{\ell\to 0}\phi_n(\bk,\ell)=0$, and
	\be
	\eta(\bk):=\frac{\ff_{02}(\bk)}{\ff_{01}(\bk)}.
	\label{eta-def}
	\ee
Taking the small-$\ell$ limit of the right-hand sides of (\ref{fz1=}) and (\ref{fz2=}), we find 
	\[\tilde\fz_n\to \frac{2\pi \eta(\bk)^{2n-3}}{\ln(k\ell)}~~~\for~~~\ell\to 0.\]

In the coincidence limit, $\ell\to 0$, the double-delta-function potential~(\ref{double-delta}) tends to the single-delta-function potential with coupling constant $\fz_1+\fz_2$, i.e., 
	\be
	v(x,y)=\fz\,\delta(x)\delta(y),\quad\quad\quad \fz:=\fz_1+\fz_2.
	\label{single-delta}
	\ee
Let $\mathring\fz$ and $\tilde\fz$ respectively denote the bare and renormalized coupling constants corresponding to $\fz$. In view of (\ref{f=1}), the scattering amplitude for (\ref{single-delta}) has the form,
	\be
	\ff(\bk',\bk)=-\sqrt{\frac{i}{8\pi k}}\:\frac{1}{\tilde\fz^{-1}+i/4}.
	\label{f=1n}
	\ee
We can use this equation together with (\ref{f=2-delta}) and (\ref{limit}) to identify the correct coincidence limit of (\ref{f=2-delta}) with
	\be
	\ff_{01}(\bk)+\ff_{02}(\bk)=\frac{1}{\tilde\fz^{-1}+i/4}.
	\label{coincidence1}
	\ee
Observe that due to the arbitrariness in the subtraction of infinities from $\mathring\fz$,
we cannot make a connection between $\tilde\fz$ and $\tilde\fz_n$. Therefore, as it stands, (\ref{coincidence1}) does not impose any constraint on $\tilde\fz_n$ or their small-$\ell$ behavior. 

The requirement that $\ff_n(\bk)$ should tend to nonzero regular functions as $\ell\to 0$ provides a simple resolution of the coincidence limit problem at the expense of making the renormalized coupling constants $\tilde\fz_n$ depend on $\ell$ and $\bk$ through Eqs.~(\ref{fz1=}) and (\ref{fz2=}). The presence of the undetermined functions $\phi_n$ and $\ff_{0n}$ in these equations shows that the standard treatment of the double-delta-function potential is not capable of describing the dependence of the scattering amplitude on the distance between the centers of the delta functions. In the remainder of this section, we describe an alternative treatment of the double-delta-function potential that is free from this short-coming.
 
Refs.~\cite{pra-2016,pra-2021} outline a dynamical formulation of stationary scattering (DFSS) whose application to the double-delta-function potential (\ref{double-delta}) does not involve divergent terms and yields (\ref{f=N}) with $A_{mn}$ given by (\ref{A=2}), \cite{jpa-2018}. This corresponds to (\ref{f=2-delta}) with $\ff_n(\bk)$ given by
	\begin{align}
	&\ff_1(\bk)= A^{-1}_{11}+A^{-1}_{12}e^{ik\ell\sin\theta_0}=
	\frac{4\big[4 \fz_2^{-1}+i-i J_0(k\ell)e^{ik\ell\sin\theta_0}\big]}{
	(4\fz_1^{-1}+i)(4\fz_2^{-1}+i)+J_0(k\ell)^2},
	\label{f1=j}\\[3pt]
	&\ff_2(\bk)= A^{-1}_{21}+A^{-1}_{22}e^{ik\ell\sin\theta_0}=	
	\frac{4\big[(4\fz_1^{-1}+i)e^{ik\ell\sin\theta_0}-i J_0(k\ell)\big]}{
	(4\fz_1^{-1}+i)(4\fz_2^{-1}+i)+J_0(k\ell)^2}.
	\label{f2=j}
	\end{align}
Because the application of DFSS to (\ref{double-delta}) does not involve any singular terms, there is no need to interpret $\fz_n$ as bare coupling constants and perform their renormalization. The fact that they determine the scattering amplitude via (\ref{f=2-delta}), (\ref{f1=j}), and (\ref{f2=j}) justifies their identification with the physical parameters of the scattering problem.

For $\ell\to 0$, $J_0(k\ell)\to 1$, and (\ref{limit}), (\ref{f1=j}), and (\ref{f2=j}) imply
	\begin{align}
	&\ff_{0n}(\bk)=\frac{4\fz_n}{4+i(\fz_1+\fz_2)}.
	\label{fnzero}
	\end{align}	
Evaluating the $\ell\to 0$ limit of (\ref{f=2-delta}) and making use of 
(\ref{limit}) and (\ref{fnzero}), we have
	\[\lim_{\ell\to 0}\ff(\bk',\bk)=-\sqrt{\frac{i}{8\pi k}}\,\frac{1}{(\fz_1+\fz_2)^{-1}+i/4}.\]
This is precisely the formula that DFSS gives for the scattering amplitude of the single-delta-function potential (\ref{single-delta}), namely (\ref{f=1-DFSS}). Therefore, DFSS not only avoids unwanted singularities of the standard treatment of the double-delta-function potential, but it also produces the correct result in the coincidence limit. 

The application of the standard treatment of multi-delta-function potentials to (\ref{double-delta}) would agree with the outcome of the DFSS provided that the renormalized coupling constants $\tilde\fz_n$ depend on $\bk$, $\ell$, and $\fz_n$ in such a way that (\ref{f1=}) and (\ref{f2=}) coincide with (\ref{f1=j}) and (\ref{f2=j}), respectively.%\footnote{The $\fz_n$ entering  (\ref{f1=j}) and (\ref{f2=j}) are finite coupling constants; they should not be confused with the bare coupling constants of Sec.~\ref{S2} which are also denoted by the same symbols.}%
~Equating the right-hand sides of these equations and solving for $\tilde\fz_n$, we find
	\bea
	\tilde\fz_1&=&\frac{4\fz_1\left\{\fz_2\left[e^{ik\ell\sin\theta_0} J_0(k\ell)-1\right]+4i\right\}}{
	\fz_1\fz_2 Y_0(k\ell)\left[J_0(k\ell)-e^{ik\ell\sin\theta_0}\right]+4\fz_2\big[e^{ik\ell\sin\theta_0}
	H^{(1)}_0(k\ell)-1\big]+16i},
	\label{fz1=2}\\[3pt]
	\tilde\fz_2&=&\frac{4\fz_2
	\left\{\fz_1\left[e^{-ik\ell\sin\theta_0} J_0(k\ell)-1\right]+4i\right\}}{
	\fz_1\fz_2 Y_0(k\ell)\big[J_0(k\ell)-e^{-ik\ell\sin\theta_0}\big]+
	4 \fz_1\big[e^{-ik\ell\sin\theta_0} H^{(1)}_0(k\ell)-1\big]+16i},
	\label{fz2=2}
	\eea
where $Y_0$ is the zero-order Bessel function of the second kind, and we have made use of the identity, $H^{(1)}_0(x)=J_0(x)+i Y_0(x)$.

Next, we calculate $\eta(\bk)$ by inserting (\ref{fnzero}) in (\ref{eta-def}). This gives
	\be
	\eta(\bk)=\frac{\fz_2}{\fz_1}.
	\label{eta=}
	\ee
It is not difficult to check that for this choice of $\eta(\bk)$, (\ref{fz1=2}) and (\ref{fz2=2}) agree with (\ref{fz1=}) and (\ref{fz2=}). Furthermore, substituting (\ref{fnzero}) in (\ref{coincidence1}), we find $\tilde\fz=\fz_1+\fz_2\neq \lim_{\ell\to 0}(\tilde\fz_1+\tilde\fz_2)=0$.

\subsection{Double-delta-function potential in 3D}

Given a double-delta-function potential in three dimensions, we can choose a coordinate system in which it takes the form
	\be
	v(x,y,z)=\fz_1\delta(x)\delta(y)\delta(z)+\fz_2\delta(x-\ell)\delta(y)\delta(z),
	\label{double-delta-3D}
	\ee
for some $\ell>0$. Comparing this relation with (\ref{multi-delta}), we see that $\bfa_1=\bzero$ and $\bfa_2=\ell\bfe_x$. Substituting these equations in (\ref{f=N-2}) and (\ref{fm=N}) and making use of (\ref{f=N-3d}) and (\ref{Amn-3d=}), we have
	\begin{align}
	&\ff(\bk',\bk)=-\frac{1}{4\pi}\,\Big[\ff_1(\bk)+\ff_2(\bk)e^{-ik\ell \sin\theta \cos\varphi}\Big],
	\label{f=2-delta-3D}\\
	&\ff_1(\bk)= A^{-1}_{11}+A^{-1}_{12}e^{ik\ell\sin\theta_0\cos\varphi_0}=
	\frac{4\pi\big[4\pi\tilde\fz_2^{-1}+ik-\ell^{-1}e^{ik\ell(\alpha_0+1)}\big]}{
	(4\pi\tilde\fz_1^{-1}+ik)(4\pi\tilde\fz_2^{-1}+ik)-\ell^{-2}e^{2ik\ell}},
	\label{f1=3D}\\[3pt]
	&\ff_2(\bk)= A^{-1}_{21}+A^{-1}_{22}e^{ik\ell\sin\theta_0\cos\varphi_0}=	
	\frac{4\pi\big[(4\pi\tilde\fz_1^{-1}+ik)e^{ik\ell\alpha_0}-\ell^{-1}e^{ik\ell}\big]}{
	(4\pi\tilde\fz_1^{-1}+ik)(4\pi\tilde\fz_2^{-1}+ik)-\ell^{-2}e^{2ik\ell}}.
	\label{f2=3D}
	\end{align}
Here we use $(k,\theta_0,\varphi_0)$ and $(k,\theta,\varphi)$ to denote the spherical coordinates of the wave vectors $\bk$ and $\bk'$, so that $\bk\cdot\bfe_x=k\sin\theta_0\cos\varphi_0$ and $\bk'\cdot\bfe_x=k\sin\theta\cos\varphi$, and $\alpha_0:=\sin\theta_0\cos\varphi_0$.

Again we demand that as $\ell\to 0$, the functions $\ff_n(\bk)$ do not diverge, i.e., there are functions $\ff_{0n}(\bk)$ fulfilling (\ref{limit}). In Appendix~B we use this requirement to derive the following analogs of (\ref{fz1=}) and (\ref{fz2=}).
	\begin{align}
	&\tilde\fz_1=\left\{
	-\frac{1}{4\pi\ell}\Big[\eta(\bk) e^{ik\ell}+ik\ell\Big]+\frac{1}{\ff_{01}(\bk)}+\phi_1(\bk,\ell)\right\}^{-1},
	\label{fz1=3D}\\
	&\tilde\fz_2=\left\{
	-\frac{1}{4\pi\ell}\Big[\eta(\bk)^{-1} e^{ik\ell}+ik\ell\Big]+\frac{1}{\ff_{02}(\bk)}+\phi_2(\bk,\ell)\right\}^{-1},
	\label{fz2=3D}
	\end{align}
where $\eta(\bk)$ is given by (\ref{eta-def}), and $\phi_n(\bk,\ell)$ are functions satisfying $\displaystyle\lim_{\ell\to 0}\phi_n(\bk,\ell)=0$. In particular, $\tilde\fz_n\to -4\pi \eta(\bk)^{2n-3}\ell$ as $\ell\to 0$.

Having established Eqs.~(\ref{fz1=3D}) and (\ref{fz2=3D}), we can state the correct coincidence limit of the scattering amplitude (\ref{f=2-delta-3D}) in the form,
	\be
	\ff_{01}(\bk)+\ff_{02}(\bk)=-\frac{1}{4\pi\tilde\fz^{-1}+ik},
	\label{coincidence1-3D}
	\ee
where $\tilde\fz$ is the renormalized coupling constant associated with $\fz:=\fz_1+\fz_2$, and we have employed (\ref{f1-3d=}). Notice that, similarly to two dimensions, $\tilde\fz\neq
\lim_{\ell\to 0}(\tilde\fz_1+\tilde\fz_2)=0$.

In Appendix~C, we use DFSS to calculate the scattering amplitude for a multi-delta-function potential in three dimensions with the centers of the delta functions located on the $x$-axis. The result is (\ref{f=N-3d}) with $d=3$ and 
	\be
	A_{mn}:=\left\{\begin{array}{ccc}
	\fz_n^{-1}+ik/4\pi &\for& m=n,\\[3pt]
	\displaystyle\frac{i\sin (|\bfa_m-\bfa_n|k)}{4\pi|\bfa_m-\bfa_n|}&\for&m\neq n.\end{array}\right.
	\label{A-DFSS-3D}
	\ee
For $N=1$, $A_{11}^{-1}=(\fz_n^{-1}+ik/4\pi)^{-1}$ and (\ref{f=N-3d}) gives (\ref{f1-3d=}) with $\tilde\fz_1$ changed to $\fz_1$. In particular, for the potential $v(\bx)=\fz\,\delta(\bx)$, we find
	\be
	\ff(\bk',\bk)=-\frac{4\pi}{4\pi\,\fz^{-1}+ik}.
	\label{f1-3d=0}
	\ee
	
For the double-delta-function potential (\ref{double-delta-3D}), we can write  (\ref{f=N-3d}) in the form (\ref{f=2-delta-3D}) with the following choices for the functions $\ff_{n}(\bk)$.
	\begin{align}
	&\ff_1(\bk)= A^{-1}_{11}+A^{-1}_{12}e^{ik\ell\alpha_0}=
	\frac{4\pi\big[4\pi \fz_2^{-1}+ik-i\ell^{-1}\sin(k\ell)e^{ik\ell\alpha_0}\big]}{
	(4\pi \fz_1^{-1}+ik)(4\pi \fz_2^{-1}+ik)+\ell^{-2}\sin^2(k\ell)},
	\label{f1=3D-DFSS}\\[3pt]
	&\ff_2(\bk)= A^{-1}_{21}+A^{-1}_{22}e^{ik\ell\alpha_0}=	
	\frac{4\pi\big[(4\pi \fz_1^{-1}+ik)e^{ik\ell\alpha_0}-i\ell^{-1}\sin(k\ell)\big]}{
	(4\pi \fz_1^{-1}+ik)(4\pi \fz_2^{-1}+ik)+\ell^{-2}\sin^2(k\ell)},
	\label{f2=3D-DFSS}
	\end{align}
where we have employed (\ref{A-DFSS-3D}). It is easy to evaluate the $\ell\to 0$ limit of the right-hand sides of (\ref{f1=3D-DFSS}) and (\ref{f2=3D-DFSS}). Inserting the result in (\ref{limit}) gives
	\be
	\ff_{0n}(\bk)=\frac{4\pi \fz_n}{4\pi+ik(\fz_1+\fz_2)}.
	\label{f-zero-3d-dfss}
	\ee
According to (\ref{f=2-delta-3D}) and (\ref{f-zero-3d-dfss}), 
	\[\lim_{\ell\to 0}\ff(\bk',\bk)=-\frac{\ff_{01}(\bk)+\ff_{02}(\bk)}{4\pi}=-\frac{4\pi}{4\pi(\fz_1+\fz_2)^{-1}+ik}.\]
Comparing this relation with (\ref{f1-3d=0}), we conclude that the application of DFSS to the double-delta-function potential (\ref{double-delta-3D}) produces a formula for the scattering amplitude that has the correct coincidence limit. 

The results obtained using the standard method based on the Lippmann-Schwinger equation agree with those of DFSS provided that we choose the renormalized coupling constants of the former approach in such a way that (\ref{f1=3D}) and (\ref{f1=3D})  coincide with (\ref{f1=3D-DFSS}) and (\ref{f1=3D-DFSS}), respectively. This condition gives rise to the following three-dimensional analogs of (\ref{fz1=2}) and (\ref{fz2=2}). 
	\bea
	\tilde\fz_1&=&\frac{4\pi\fz_1\big\{\fz_2\ell^{-1}[k\ell-e^{ik\ell\alpha_0}\sin(k\ell)]-4\pi i\big\}}{
	\fz_1\fz_2\ell^{-2}\cos(k\ell)[\sin(k\ell)-k\ell\,e^{ik\ell\alpha_0}]
	+4\pi\fz_2\ell^{-1}[k\ell+i e^{ik\ell(\alpha_0+1)}]-16\pi^2i},\nn\\
	\tilde\fz_2&=&\frac{4\pi\fz_2\big\{\fz_1\ell^{-1}[k\ell-e^{-ik\ell\alpha_0}\sin(k\ell)]-4\pi i\big\}}{
	\fz_1\fz_2\ell^{-2}\cos(k\ell)[\sin(k\ell)-k\ell\,e^{-ik\ell\alpha_0}]
	+4\pi\fz_1\ell^{-1}[k\ell\,+i e^{ik\ell(-\alpha_0+1)}]-16\pi^2i}.\nn
	\eea
The small-$\ell$ behavior of these relations are described by (\ref{fz1=3D}) and (\ref{fz2=3D}) with $\ff_{0n}(\bk)$ and $\eta(\bk)$ given by (\ref{f-zero-3d-dfss}) and (\ref{eta=}), respectively.

\section{Concluding remarks}
\label{S4}

The emergence of unwanted singularities in dealing with physics problems has provided the much needed clues for making important developments in theoretical physics. Among the most notable of these is the development of renormalization schemes which have proven to be indispensable in the study of fundamental interactions. Delta-function potentials in two and three dimensions provide simple exactly solvable non-relativistic toy models whose standard treatment requires a coupling constant renormalization. This has made these potentials an ideal pedagogical tool for teaching the basic idea and methods of the renormalization program. During the past four decades, there have appeared many publications on this subject \cite{thorn,jackiw,mead,manuel,Adhikari1,Adhikari2,Mitra,Rajeev-1999,Nyeo,Camblong,teo-2006,teo-2010,teo-2013,ferkous,ap-2019}, but none pay attention to the coincidence-limit problem for multi-delta-function potentials. 

Multi-delta-function potentials model a collection of point scatterers whose sizes are much smaller than the wavelength of the incident wave. These scatterers have nevertheless nonzero spatial extensions. Therefore they can never coincide. One can use this argument to question the physical relevance of the coincidence-limit problem. But it cannot explain how shrinking the distance between two of the point scatterers can nullify the scattering effects of the others whose positions remain unchanged.

In this article we draw attention to the fact that the application of an alternative approach to scattering theory, namely the dynamical formulation of stationary scattering (DFSS), produces an expression for the scattering amplitude of multi-delta-function potentials in two and three dimensions that does not suffer from the coincidence-limit problem. Our attempts at exploring the relationship between the outcomes of DFSS and the standard approach of using the Lippmann-Schwinger equation have led us to realize that the renormalized coupling constants $\tilde\fz_n$ appearing in the latter approach depend not only on the energy scale of the problem (determined by the wavenumber $k$ of the incident wave) but other relevant physical parameters such as the distances between the point scatterers. More importantly, our results show that the standard treatment of multi-delta-function potentials in two and three dimensions is not capable of determining the dependence of $\tilde\fz_n$ on these parameters. The requirement that the outcome must have a consistent coincidence limit provides some information about the behavior of $\tilde\fz_n$ when the distance(s) between two or more of point scatterers become much smaller than $k^{-1}$, but it does not fix the functional form of $\tilde\fz_n$. This in turn implies that the formula we obtain using the standard approach for the scattering amplitude does not describe its dependence on the location of the point scatterers. This is in sharp contrast to the formula we obtain using DFSS.

\section*{Acknowledgements}
This work has been supported by the Scientific and Technological Research Council of Turkey (T\"UB\.{I}TAK) in the framework of the project 120F061 and by Turkish Academy of Sciences (T\"UBA).

\section*{Appendix~A: Coincidence limit of multi-delta-function potentials in DFSS}

Eq.~(\ref{f=N-2}) gives the scattering amplitude for multi-delta-function potential (\ref{multi-delta}) provided that (\ref{fm=N}) holds. This means that the following equation holds for all $m\in\{1,2\cdots,N\}$.
	\be
	\sum_{n=1}^NA_{mn}\ff_n(\bk)=e^{i\bfa_m\cdot\bk}.
	\label{fm=N-appA}
	\ee
As shown in Ref.~\cite{jpa-2018} for two dimensions and in Appendix~C for three dimensions, when the centers of the delta-functions contributing to (\ref{multi-delta}) lie on a straight line, we can apply DFSS to determine $A_{mn}$. This leads to (\ref{A=2}) and (\ref{A-DFSS-3D}) for two and three dimensions, respectively. We can express these equations in the following unified form.
	\be
	A_{mn}=\left\{\begin{array}{ccc}
	\fz_n^{-1}+\gamma_d & \for & m=n,\\[3pt]
	\gamma_d\, Q_d(k|\bfa_m-\bfa_n|)&\for&m\neq n,\end{array}\right.
	\label{Amn=110}
	\ee
where
	\begin{align}
	&\gamma_d:=\left\{\begin{array}{ccc}
	\frac{i}{4} & \for & d=2,\\[3pt]
	\frac{ik}{4\pi}&\for&d=3,\end{array}\right.
	&&Q_d(x):=\left\{\begin{array}{ccc}
	J_0(x) & \for & d=2,\\[3pt]
	\displaystyle\frac{\sin x}{x}&\for&d=3.\end{array}\right.\nn
	\end{align}
Notice that for both $d=2$ and $d=3$,
	\be
	\lim_{x\to 0} Q_d(x)=1.
	\label{Q-lim}
	\ee

Suppose that the distances between the centers of $\nu$ of the delta functions contributing to (\ref{multi-delta}) tend to zero. We refer to these as the ``merging delta functions.'' By relabeling the centers and coupling constants $\fz_i$ of the delta functions appearing in (\ref{multi-delta}), we can assume without loss of generality that the merging delta functions are labelled by $N-\nu+1, N-\nu+2,\cdots,N$. We wish to explore the behavior of the scattering amplitude $\ff(\bk',\bk)$ for the potential (\ref{multi-delta}) in
the coincidence limit,
	\be
	|\bfa_n-\bfa_{N-\nu+1}|\to 0~~~\for~~~n\in\{N-\nu+2,N-\nu+3,\cdots,N\}.
	\label{C-lim}
	\ee
This implies
	\be
	v(\bx)\to v_\star(\bx):=\sum_{n=1}^{N-\nu}\fz_n\delta(\bx-\bfa_n)+
	\fz_\star\,\delta(\bx-\bfa_{N-\nu+1}),
	\label{v-star}
	\ee
where $\fz_\star:=\fz_{N-\nu+1}+\fz_{N-\nu+2}+\cdots+\fz_N$. $v_\star$ is a (multi-)delta-function potential consisting of $N-\nu+1$ delta functions. We wish to show that in the coincidence limit given by (\ref{C-lim}), $\ff(\bk,\bk')$ tends to the scattering amplitude of $v_\star$. It is not difficult to see that if we can prove this assertion for $\nu=2$, it will hold for $\nu>3$. This is simply because we can achieve (\ref{C-lim}) by making pairs of delta functions merge one at a time. Therefore, we confine our attention to the case $\nu=2$ where
	\be
	\fz_\star=\fz_{N-1}+\fz_N,
	\label{z-star}
	\ee
and (\ref{C-lim}) means
	\be
	\bfa_N\to\bfa_{N-1}.
	\label{C-lim-2}
	\ee
In this limit,
	\begin{align}
	&\ff(\bk',\bk)\to \frac{c_d}{\sqrt{k^{3-d}}}\left\{\sum_{n=1}^{N-2}\ff_n(\bk) e^{-i\bfa_n\cdot\bk'}+[\ff_{N-1}(\bk)+\ff_N(\bk)]e^{-i\bfa_{N-1}\cdot\bk'}\right\},
	\label{u111}\\
	&A_{N-1\, m}=A_{m\,N-1}\to\left\{\begin{array}{ccc}
	\gamma_d\, Q_d(|\bfa_m-\bfa_{N-1}|)&\for&m<N-1,\\
	\fz_{N-1}^{-1}+\gamma_d &\for & m=N-1,\\
	\gamma_d &\for& m=N,\end{array}\right.	
	\label{u112}\\
	&A_{N\,m}=A_{m\,N}\to\left\{\begin{array}{ccc}
	\gamma_d\, Q_d(|\bfa_m-\bfa_{N-1}|)&\for&m<N-1,\\
	\gamma_d &\for & m=N-1,\\
	\fz_{N}^{-1}+\gamma_d  &\for& m=N,\end{array}\right.
	\label{u113}
	\end{align}
where we have employed (\ref{f=N-2}), (\ref{Amn=110}), and (\ref{Q-lim}). We also note that (\ref{C-lim-2}) does not affect $A_{mn}$ for $m<N-1$ and $n<N-1$; they are still given by (\ref{Amn=110}). 

Next, we examine the effect of (\ref{C-lim-2}) on $\ff_m(\bk)$. We can use (\ref{fm=N-appA}), (\ref{u112}), and (\ref{u113}) to show that, in this limit, (\ref{fm=N-appA}) gives
	\begin{align}
	&\sum_{n=1}^{N-2}A_{m\, n}\,\ff_n(\bk)+\gamma_d Q_d(|\bfa_m-\bfa_{N-1}|)[\ff_{N-1}(\bk)+\ff_N(\bk)]=e^{i\bfa_m\cdot\bk}~~\for~~m<N-1,
	\label{u114}\\
	&\sum_{n=1}^{N-2}A_{N-1\, n}\,\ff_n(\bk)+(\fz_{N-1}^{-1}+\gamma_d)\ff_{N-1}(\bk)+
	\gamma_d\,\ff_N(\bk)=e^{i\bfa_{N-1}\cdot\bk},
	\label{u115}\\
	&\sum_{n=1}^{N-2}A_{N n}\,\ff_n(\bk)+\gamma_d\,\ff_{N-1}(\bk)+
	(\fz_{N}^{-1}+\gamma_d)\ff_N(\bk)=e^{i\bfa_{N-1}\cdot\bk}.
	\label{u116}
	\end{align}
According to (\ref{u112}) and (\ref{u113}), $A_{N\, n}=A_{N-1\, n}$ for $n<N-1$. This observation together with (\ref{u115}) and (\ref{u116}) imply that $\ff_N(\bk)=\fz_{N-1}^{-1}\,\fz_N\,\ff_{N-1}(\bk)$. With the help of this equation and (\ref{z-star}), we can write (\ref{u115}) as
	\be
	\sum_{n=1}^{N-2}A_{N-1\, n}\,\ff_n(\bk)+(\fz_\star^{-1}+\gamma_d)[\ff_{N-1}(\bk)+\ff_N(\bk)]=e^{i\bfa_{N-1}\cdot\bk}.
	\label{u118}
	\ee
	
Let us introduce,
	\begin{align}
	&\check\fz:=\left\{\begin{array}{ccc}
	\fz&\for&m<N-1,\\
	\fz_\star&\for&m=N-1,\end{array}\right.
	&&
	\check\ff_m(\bk):=\left\{\begin{array}{ccc}
	\ff_m(\bk)&\for&m<N-1,\\
	\ff_{N-1}(\bk)+\ff_N(\bk)&\for&m=N-1.\end{array}\right.
	\end{align}
Then (\ref{u111}), (\ref{u114}), and (\ref{u118}) show that in the coincidence limit the scattering amplitude is given by
	\be
	\frac{c_d}{\sqrt{k^{3-d}}}\sum_{n=1}^{N-1}\check\ff_n(\bk) e^{-i\bfa_n\cdot\bk'},
	\label{u118n}
	\ee
with $\check\ff_m(\bk)$ satisfying $\sum_{n=1}^{N-1}\check A_{mn}\check\ff_m(\bk)=e^{i\bfa_m\cdot\bk}$, and
	\[\check\bA_{mn}:=\left\{\begin{array}{ccc}
	\check\fz_n^{-1}+\gamma_d &\for&m=n,\\
	\gamma_d\,Q(|\bfa_m-\bfa_n|)&\for&m\neq n.\end{array}\right.\]
Comparing this relation with (\ref{Amn=110}), we identify (\ref{u118n}) with the scattering amplitude for the potential $v_\star$. This concludes the proof that the scattering amplitude given by (\ref{f=N-2}), (\ref{fm=N}), and (\ref{Amn=110}) has a consistent coincidence limit. We have been unable to extend this result to situations where the centers of the delta functions contributing to the potential do not lie on a line, simply because the application of DFSS to these potentials leads to technical difficulties.

\section*{Appendix~B: Derivation of (\ref{fz1=}), (\ref{fz2=}), (\ref{fz1=3D}), and (\ref{fz2=3D})}

Eqs.~(\ref{fz1=}) and (\ref{fz2=}) reveal the small-$\ell$ behavior of the renormalized coupling constants $\tilde\fz_n$. To derive these equations, first we introduce,
	\begin{align}
	&h:=\frac{i}{4} H_0^{(1)}(k\ell),
	&&\xi_n:=\tilde\fz_n^{-1}-h+\frac{i}{4},
	\label{app1}\\
	&\mu_1:=-\big(e^{ik\ell\sin\theta_0}-1\big)h,
	&&\mu_2:=\big(e^{ik\ell\sin\theta_0}-1\big)(\xi_1+h),
	\label{app1n}
	\end{align}
and use them to express (\ref{f1=}) and (\ref{f2=}) in the form,
	\begin{align}
	&\ff_1(\bk)=\frac{\xi_2+\mu_1}{\xi_1\xi_2+(\xi_1+\xi_2)h},
	&&\ff_2(\bk)=\frac{\xi_1+\mu_2}{\xi_1\xi_2+(\xi_1+\xi_2)h}.
	\label{fffs}
	\end{align}
According to (\ref{H-asymp}) and (\ref{app1}),  $h$ diverges logarithmically as $\ell\to 0$. This together with (\ref{app1n}) imply
	\begin{align}
	&\lim_{\ell\to 0}\mu_1=0,
	&&\lim_{\ell\to 0}(\xi_1+\mu_2)=\lim_{\ell\to 0}\xi_1.
	\label{app02}
	\end{align}
With the aid of these relations, we can use (\ref{limit}) and (\ref{fffs}) to show that
	\begin{align}
	&\ff_{01}(\bk)=\lim_{\ell\to 0}\frac{1}{\xi_1+[\eta(\bk)+1]h},
	&&\ff_{02}(\bk)=\lim_{\ell\to 0}\frac{1}{\xi_2+[\eta(\bk)^{-1}+1]h},
	\label{app3}
	\end{align}
and 
	\be
	\eta(\bk)=\lim_{\ell\to 0}\frac{\xi_1}{\xi_2}.
	\label{limit-2}
	\ee	
Next, we introduce
	\begin{align}
	&\phi_1(\bk,\ell):=\xi_1+[\eta(\bk)+1]h-\frac{1}{\ff_{01}(\bk)},
	&&\phi_2(\bk,\ell):=\xi_2+[\eta(\bk)^{-1}+1]h-\frac{1}{\ff_{02}(\bk)}.
	\label{app4}
	\end{align}
These functions satisfy, $\displaystyle\lim_{\ell\to 0}\phi_n(\bk,\ell)=0$, by virtue of (\ref{app3}). Substituting the second relation in (\ref{app1}) in (\ref{app4}) and solving the resulting equations for $\tilde\fz_n$, we arrive at (\ref{fz1=}) and (\ref{fz2=}).

We can similarly derive (\ref{fz1=3D}) and (\ref{fz2=3D}). To see this, first we note that we can express (\ref{f1=3D}) and (\ref{f2=3D}) in the form (\ref{fffs}) provided that we redefine $h$, $\xi_n$, and $\mu_n$ as follows.
	\begin{align}
	&h:=\frac{e^{ik\ell}}{4\pi\ell},
	&&\xi_n:=\tilde\fz_n^{-1}-h+\frac{ik}{4\pi},
	\label{app1-3D}\\
	&\mu_1:=-\big(e^{ik\ell\alpha_0}-1\big)h,
	&&\mu_2:=\big(e^{ik\ell\alpha_0}-1\big)\big(\xi_1+h\big).
	\label{app1-3Dn}
	\end{align}
If $\xi_1$ or $\xi_2$ tends to a finite limit as $\ell\to 0$,  (\ref{app1-3D}) and (\ref{app1-3Dn}) imply that the right-hand side of both of the equations in (\ref{fffs}) tend to zero as $\ell\to 0$. This contradicts the requirement that $\ff_{0n}(\bk)$ are nonzero functions. Therefore $\xi_1$ and $\xi_2$ must both tend to infinity as $\ell\to 0$. We can use this observation together with (\ref{limit}) and (\ref{fffs}) to show that (\ref{limit-2}) holds. Using this equation together with (\ref{limit}) and (\ref{fffs}), we are led to (\ref{app3}). This in turn shows that the functions $\phi_n(\bk,\ell)$ given by (\ref{app4}) fulfill $\displaystyle\lim_{\ell\to 0}\phi_n(\bk,\ell)=0$. Eqs.~(\ref{fz1=3D}) and (\ref{fz2=3D}) follow from (\ref{app4}) and (\ref{app1-3D}).

\section*{Appendix~C: Application of DFSS to multi-delta-function potentials in 3D}

Consider a multi-delta-function potential~(\ref{multi-delta}) in three dimensions and suppose that the centers of the delta functions are on a straight line that we identify with the $x$ axis, i.e., $\bfa_n=a_n\bfe_x$. Then the potential has the form 
	\be
	v(x,y,z)=g(x,y)\delta(z),
	\label{app-100}
	\ee
where
	\be
	g(x,y)=\delta(y)\sum_{n=1}^N\fz_n\delta(x-a_n),
	\label{g=}
	\ee
and we can use the results of Sec.~8 of Ref.~\cite{pra-2021} to compute the scattering amplitude. In the following we give the details of this calculation.

First, we introduce a suitable notation. Given $\bu\in\R^3$, we denote the projection of $\bu$ onto the $x$-$y$ plane by $\vec u$; if in Cartesian coordinates $\bu=(u_x,u_y,u_z)$, then $\vec u=(u_x,u_y)$. In particular, $\vec x=(x,y)$ because $\bx=(x,y,z)$. We also use the hybrid notation: $\bu=(\vec u,u_z)$.

Now, consider a scattering setup where the source of the incident wave is located at $z=-\infty$ or $z=+\infty$, then every solution of the Schr\"odinger equation (\ref{sch-eq}) for a short-range potential $v$ satisfies
    \be
    \psi(\vec x,z)\to\int_{\sD_k}\frac{d^2\vec p}{4\pi^2\varpi(\vec p)}\: e^{i\vec p\cdot\vec x}
    \left[\breve A_\pm(\vec p)e^{i\varpi(\vec p)z}+\breve B_\pm(\vec p)e^{-i\varpi(\vec p)z}\right]~~\for~~z\to\pm\infty,
    \nn
    \ee
where
    \begin{align}
    &\sD_k:=\left\{\vec p\in\R^2~|~|\vec p|<k~\right\},
    &&\varpi(\vec p):=\left\{\begin{array}{ccc}
    \sqrt{k^2-|\vec p|^2}&\for&|\vec p|<k,\\[3pt]
    i\sqrt{|\vec p|^2-k^2}&\for&|\vec p|\geq k,\end{array}\right.
    \label{sD=}
    \end{align}
and $\breve A_\pm$ and $\breve B_\pm$ are functions of $\vec p\in\R^2$ that vanish for $|\vec p|\geq k$. We denote the set of functions with this property by $\sF_k$, so that $\breve A_\pm,\breve B_\pm\in\sF_k$. It is not difficult to see that the scattering amplitude of the potential should be related to $\breve A_\pm$ and $\breve B_\pm$. If the source of the incident wave is at $z=+\infty$, we have \cite{pra-2021},
	\begin{align}
	&\breve A_-=0,\quad\quad\quad\quad\quad\quad
	\breve B_+=4\pi^2\varpi(\vec k)\,\delta_{\vec k},
	\label{app-A-101}\\
	&\ff(\bk',\bk)=-\frac{i}{2\pi}\times\left\{
    	\begin{array}{ccc}
    	\breve A_+(\vec k')&\for&\vartheta\in[0,\frac{\pi}{2}),\\
    	\breve B_-(\vec k')-4\pi^2\varpi(\vec k)\delta(\vec k'-\vec k)&\for&\vartheta\in(\frac{\pi}{2},\pi],\end{array}\right.
    	\label{app-A-102}
    	\end{align}
where $(k,\vartheta_0,\varphi_0)$ and $(k,\vartheta,\varphi)$ are respectively the spherical coordinates of $\bk$ and $\bk'$, so that
	\begin{align}
	&\vec k:=k\sin\vartheta_0(\cos\varphi_0\,\bfe_x+\sin\varphi_0\,\bfe_y),
	&&\vec k':=k\sin\vartheta(\cos\varphi\,\bfe_x+\sin\varphi\,\bfe_y),\nn
 	\end{align}
and $\delta_{\vec k}$ is the delta function in two dimensions that is centered at $\vec k$, i.e., $\delta_{\vec k}(\vec p):=\delta(\vec p-\vec k)$. Notice that $|\vec k|=k\sin\vartheta_0$, $|\vec k'|=k\sin\theta$, and $\varpi(\vec k)=k|\cos\vartheta|$.
	
The fundamental transfer matrix $\widehat{\breve\bM}$ is a linear operator acting in the space $\sF_k^2$. It is conveniently expressed as the $2\times 2$ matrix with operator entries $\widehat{\breve M}_{ij}:\sF_k\to\sF_k$ that fulfills
	\[\widehat{\breve\bM}\left[\begin{array}{c}
	\breve A_-\\
	\breve B_-\end{array}\right]=\left[\begin{array}{c}
	\breve A_+\\
	\breve B_+\end{array}\right].\]
By virtue of this relation and (\ref{app-A-101}),  we have
	\begin{align}
	 &\breve A_+=\widehat{\breve M}_{12}\breve B_-,
    	\label{z203b-3d}\\
    	&\widehat{\breve M}_{22}\breve B_-=4\pi^2\varpi(\vec k)\,\delta_{\vec k}.
    	\label{z206b-3d}
	\end{align}

In Ref.~\cite{pra-2021}, we calculate the fundamental transfer matrix for potentials of the form (\ref{app-100}) and show that 			\begin{align}
	&\big(\widehat{\breve M}_{12}\phi\big)(\vec p)=-\frac{i}{8\pi^2}\int_{\sD_k} 
	\frac{\tilde{\tilde g}(\vec p-\vec q)\phi(\vec q)}{\varpi(\vec q)}\,d^2\vec q,
	\label{M12}\\
	&\big(\widehat{\breve M}_{22}\phi\big)(\vec p)=\phi(\vec p)-\big(\widehat{\breve M}_{12}\phi\big)(\vec p),
	\label{M22}
	\end{align}
where $\phi\in\sF_k$,  and $\tilde{\tilde g}$ stands for the two-dimensional Fourier transform of $g$ which has the form
	\be
	\tilde{\tilde g}(\vec p):=\int_{\R^2}d^2\vec x\:e^{-i\vec p\cdot\vec x}g(\vec x)=
	\sum_{n=1}^N\fz_n\,e^{-i\vec a_n\cdot\vec p}.
	\nn%\label{app-103}
	\ee
Substituting the last equation in (\ref{M12}) and making use of (\ref{sD=}) and (\ref{z203b-3d}) , we obtain
	\begin{align}
	&\breve A_+(\vec p)=	\big(\widehat{\breve M}_{12}\breve B_-\big)(\vec p)
	=-\frac{i}{2}\sum_{n=1}^N \fz_n\overline{\breve B}_-(\vec a_n) e^{-i\vec a_n\cdot\vec p},
	\label{app-104}
	\end{align}
where
	\be
	\overline{\breve B}_-(\vec x):=\frac{1}{4\pi^2} \int_{\sD_k}\frac{e^{i\vec q\cdot\vec x}\breve B_-(\vec q)}{\sqrt{k^2-|\vec q|^2}}\,d^2\vec q.
	\label{app-105}
	\ee
	
Next, we use (\ref{M22}) and (\ref{app-104}) to express (\ref{z206b-3d}) in the form,
	\be
	\breve B_-(\vec p)=-\frac{i}{2}\sum_{n=1}^N\fz_n\overline{\breve B}_-(\vec a_n)
	e^{-i\vec a_n\cdot\vec p}+4\pi^2\varpi(\vec k)\delta(\vec p-\vec k).
	\label{app-106}
	\ee
Substituting this equation in the right-hand side of (\ref{app-105}) and setting $\vec x=\vec a_m$, we find the following system of linear equations for $X_n:=\fz_n\overline{\breve B}_-(\vec a_n)$.
	\be
	\sum_{n=1}^N A_{mn}X_n=e^{i\bk\cdot \bfa_m},
	\label{app-107}
	\ee
where $A_{mn}$ are given by (\ref{A-DFSS-3D}),
and we have made use of $\vec k\cdot \vec a_m=\bk\cdot \bfa_m$, $|\vec a_m-\vec a_n|=|a_m-a_n|=|\bfa_m-\bfa_n|$, and 
	\[\int_{\sD_k}\frac{e^{i\vec a\cdot\vec q}}{\sqrt{k^2-|\vec q|^2}}\,d^2\vec q=\frac{2\pi\sin(|\vec a| k)}{|\vec a|}.\]
	
Assuming that $\bA$ is invertible, which happens when there are no spectral singularities \cite{prl-2009}, we can express the solution of (\ref{app-106}) in terms of the entries $A_{mn}^{-1}$ of $\bA^{-1}$. This allows us to determine $\overline{\breve B}_-(\vec a_n)=\fz_n^{-1}X_n$. Substituting the result in (\ref{app-104}) and (\ref{app-106}), we find 
	\[{\breve A}_+(\vec p)={\breve B}_-(\vec p)-4\pi^2\varpi(\vec k)\delta(\vec p-\vec k)=
	-\frac{i}{2}\sum_{m,n=1}^N A^{-1}_{mn} e^{i(\bfa_n\cdot\bk-\bfa_m\cdot\bp)}.\]
Using this relation in (\ref{app-A-102}), we recover the formula (\ref{f=N-3d}) for the scattering amplitude with $d=3$. 
Because the multi-delta-function potential (\ref{app-100}) is invariant under a reflection about the $x$-$y$ plane, this formula holds also for situations where the source of the incident wave is located at $z=-\infty$.

\ed

In the coincidence limit, where $\ell\to 0$, Eq.~(\ref{f=N-2}) should reproduce the scattering amplitude for a single delta function with center $\bzero$ and renormalized coupling constant:
	\be
	\tilde\fz:=\lim_{\ell\to 0}\left(\tilde\fz_1+\tilde\fz_2\right).
	\label{tilde-fz}
	\ee
In view of (\ref{f=1}), this implies that 
	\be
	\ff_1(\bk)+\ff_2(\bk)\to(\tilde\fz^{-1}+i/4)^{-1}$. With the help of (\ref{H-asymp}) and (\ref{f1=}) -- (\ref{tilde-fz}) we can express this condition in the form
	\be
	\lim_{\ell\to 0}\,\left[\tilde\fz_1+\tilde\fz_2+\frac{8i}{H_0^{(1)}(k\ell)-1}\right]=0.
	\label{condi}
	\ee	
	\be
	\lim_{\ell\to 0}\,\left[\tilde\fz_1+\tilde\fz_2+\frac{4\pi}{\ln(k\ell)}\right]=0.
	\label{condi}
	\ee